# Atomismo y Mecánica Cuántica

Enric Pérez Canals
Blai Pié Valls

Universitat de Barcelona
Departament de Física Fonamental

Resumen

Nos proponemos discutir, desde un punto de vista historiográfico, cuál fue el grado de certeza que los físicos involucrados directamente en el nacimiento de la Mecánica Cuántica (Heisenberg, Born, Jordan, Dirac y Schrödinger) otorgaban a la hipótesis atomista, a partir de 1925, y hasta la celebración del quinto Congreso Solvay, en octubre de 1927. En particular, seguiremos el tratamiento que se hizo de la partícula libre, y cómo éste influyó de forma determinante en las interpretaciones que surgieron de la nueva mecánica. Mostramos que, a pesar de centrarnos en un lapso de tiempo tan relativamente corto y pegado a los años fundacionales, prácticamente todos los problemas y discusiones que aún hoy siguen vivos ya se plantearon entonces. Salvo en el caso de Schrödinger, apenas hubo cambios de valoración respecto a la hipótesis atomista.

Palabras clave: Mecánica Cuántica, Atomismo, Partícula libre, Identidad, Dualidad onda-corpúsculo.

Abstract. Atomism and Quantum Mechanics.

We discuss, from a historiographical point of view, which was the degree of certainty that the physicists directly involved in the birth of Quantum Mechanics (Heisenberg, Born, Jordan, Dirac and Schrödinger) gave to the atomistic hypothesis, starting from 1925, and until the celebration of the 5th Solvay Conference, in October 1927. In particular, we analyze how was tackled the problem of the free particle, and how that affected decisively the interpretations proposed for the new mechanics. We show that, despite focusing on such a narrow period of time, so close to those seminal years, practically all the questions which still prevail nowadays, were then already posed. Except in the case of Schrödinger, there hardly were changes of appreciation of the atomistic hypothesis.

> "… we cannot really alter our manner of thinking in space and time, and what we cannot comprehend within it we cannot understand at all. There *are* such things- but I do not believe that atomic structure is one of them."
>
> E. Schrödinger, 1926

En un congreso de ontología la pregunta '¿qué es la materia?' suena pertinente. Nosotros formularemos una cuestión íntimamente relacionada, menos interesante, pero, como veremos, no menos controvertida: según la Mecánica Cuántica ¿es la materia infinitamente divisible? Y no lo haremos desde un punto de vista científico o filosófico, sino historiográfico. Esto es, nos preguntaremos si los fundadores de la Mecánica Cuántica concluyeron que había una suerte de partículas elementales o, por el contrario, se plantearon que no, que los desvelos cuánticos habían puesto en entredicho el sustrato atomista de la realidad.

La pertinencia de nuestra pregunta está justificada por el mero hecho de que en Mecánica Cuántica, por ejemplo, las partículas libres pueden representarse como ondas planas, y que las partículas confinadas en una caja se describen mediante ondas estacionarias. *Grosso modo*, es sabido que es posible localizar exactamente una partícula, pero a cambio de perder toda la información sobre su momento (*delta de Dirac*). Y del revés, un conocimiento preciso de su momento nos priva completamente de su localización (*onda plana*). Si eso son meras representaciones o descripciones es lo que discutiremos en este artículo.

Si bien la interpretación que marca la ortodoxia -la llamada interpretación de Copenhague- establece que ninguna de las dos imágenes, corpuscular u ondulatoria, hace justicia a lo que pasa realmente, y prescribe tomar dichas imágenes como meras aproximaciones o recursos para salir del paso, la presencia y proliferación de explicaciones basadas en electrones que colisionan, protones que salen despedidos, etc. es frecuente y hasta mayoritaria, no sólo en textos divulgativos sino también en especializados.

No faltan, todo hay que decirlo, voces que denuncian dichos usos,[1] pero, por poner sólo un caso, el anuncio a bombo y platillo en los medios, el pasado verano, del descubrimiento de la 'partícula de Higgs' ilustra hasta qué punto los hallazgos de la era fundacional de la Mecánica Cuántica no han calado ni entre divulgadores ni entre expertos. El abuso de lenguaje que se comete al llamar átomo a una porción de materia divisible se queda corto al llamar partícula a algo que es directamente inconcebible.

Nosotros nos hemos propuesto analizar los escritos con que se estructuró inicialmente la Mecánica Cuántica, para ver hasta qué punto la confusión reinante proviene del origen o el asunto se fue enmarañando con el paso de los años. Para ello nos hemos limitado al periodo que va desde la aparición del primer artículo de Heisenberg, en setiembre de 1925, hasta la celebración del 5º Congreso Solvay, en octubre de 1927. Dejaremos para otra ocasión el análisis de los primeros intentos de construir una teoría cuántica de campos.

Por razones de espacio, nos dedicaremos principalmente a los fundadores, desatendiendo otros físicos cuya aportación fue relevante, pero que no firmaron los artículos en los que se pusieron las bases de la nueva teoría. Esto es, nos centramos en Heisenberg, Born, Jordan, Dirac y Schrödinger. Aunque incidentalmente, también nos referiremos a Pauli, De Broglie, Bohr y Einstein.

## 1. Atomismo decimonónico y cuantos de energía

Teorías, filosofías, hipótesis atomistas y continuistas han convivido en muchas ocasiones a lo largo de la historia de Occidente. Ejemplos notables de ello son las cosmovisiones epicúreas y estoicas, o las discusiones en torno a la naturaleza de la luz, con Newton como abanderado de su constitución corpuscular y Huygens de la ondulatoria.

En libros de Física o Historia de la Física suele darse por hecho que esa dicotomía dejó de tener sentido allá por la segunda década del siglo XX, cuando el atomismo pasó de ser una postura filosófica a convertirse en una hipótesis científica demostrada.

Para citar algunos de los autores que contribuyeron a cientificizar la vieja idea de la imposibilidad de dividir infinitamente la materia, hay que remontarse a los fundadores de la teoría cinética. Los Herapath, Waterston o el mismo Clausius iniciaron el estudio de los procesos térmicos a partir de las propiedades mecánicas de sus constituyentes últimos.[2] No en vano la negación de la sustancialidad del calor y su asimilación al movimiento de los primordios tuvo mucho que ver con el resurgir del atomismo. Maxwell y Boltzmann fueron los que acabaron de culminar esa etapa inicial introduciendo el cálculo de probabilidades en la Física, e interpretando el segundo principio de la termodinámica en términos mecánicos y probabilísticos. A todo ello hay que añadir el libro de Gibbs "Elementary Principles of Statistical Mechanics", de 1902, así como los trabajos de Einstein de esos años.

---
[1] Por ejemplo : DAVIES (1984), LAUGHLIN (2010), HOLLOWOOD (2013).
[2] Véase BRUSH (1976).

Esta vía proporcionó resultados notables en la física de los procesos térmicos (en la que prácticamente todos los autores mencionados eran expertos), y se convirtió en una subdisciplina más: la Mecánica Estadística. Suele ponerse el estudio del movimiento browniano, por parte de Einstein y von Smoluchowsky (de manera independiente) como la prueba definitiva que acabó de convencer a los últimos reticentes. La prueba consistió, básicamente, en una buena correspondencia de los cálculos del movimiento medio de las partículas en suspensión, hechos a partir de presupuestos aleatorios propios de la nueva subdisciplina, y lo medido en el laboratorio por el francés Perrin, en 1910.

Pero cuando esto ocurrió, ya se había abierto una segunda ruta.[3] El desarrollo de la electrotecnia permitió, en el último tercio del siglo XIX, crear y experimentar con una serie de radiaciones desconocidas que contribuyeron muy mucho a escudriñar las intimidades de la materia. Los rayos X son de la última década del siglo, y el hito quizá más relevante a este respecto sea el electrón y su identificación con los rayos catódicos, por parte de J.J. Thomson, en 1897. De modo que el átomo dejó de ser, en poco tiempo, indivisible. A estos nuevos fenómenos se sumaron las radiaciones $\alpha$, $\beta$ y otras.

Estudios hechos a partir de estos métodos de exploración, llevados a cabo (pero no sólo) en Inglaterra, contribuyeron a establecer el atomismo y a promover incluso los primeros modelos atómicos. El de Thomson (*plum cake*) data de 1903. De Manchester surgiría, en 1913, el modelo de Bohr, inspirado en uno anterior de Rutherford, de 1911.

Por entonces la teoría cuántica ya llevaba unos años en danza, aunque sin tener unos límites demasiado definidos.[4] En 1900, Planck cuantizó la energía de los resonadores de la cavidad radiante, pero sin darle trascendencia a un paso que luego se vio como revolucionario. Fue más de cinco años después cuando Einstein y Ehrenfest, independientemente, empezaron a hacer notar la peculiaridad y relevancia de la hipótesis planckiana. En los años sucesivos la comunidad de físicos fue tomando conciencia de la problemática que planteaba el estudio del mundo atómico, hasta que en 1911, en el Primer Congreso Solvay, se certificó la existencia de un grave problema en el corazón de la Física: aplicada al mundo microscópico conducía a contradicciones insalvables.

De modo que eso ocurrió prácticamente al mismo tiempo que la consolidación del atomismo. A partir de entonces, la teoría cuántica vivió años de tentativas y de desarrollo que dieron, a principios de los años 20, en una crisis, en un fracaso manifiesto de los métodos empleados hasta entonces. Éstos, consistían básicamente en hacer uso de la mecánica y electrodinámica ordinarias, pero con algunos parches, algunos remiendos. Alrededor de 1924, para algunos se había hecho ya evidente que para dar cuenta del comportamiento de la materia a nivel microscópico habría que empezar de cero y plantearse incluso renunciar a la visualización de los procesos elementales y la causalidad. El mundo microscópico no era un trasunto del macroscópico.

¿Afectó eso al establecimiento del atomismo? ¿Se llegó a poner en duda una hipótesis tan recientemente adoptada?

**2. La época fundacional**

Heisenberg fue el primero en articular una propuesta.[5] Planteó una reinterpretación de las ecuaciones de la mecánica ordinaria. Envió su artículo a publicar a finales de julio de 1925, y éste apareció publicado en el mes de setiembre. Poco después, le siguieron un artículo donde Born y

---

[3] Véase PAIS (1986).
[4] Véase SÁNCHEZ RON (2001).
[5] HEISENBERG (1925).

Jordan identificaban el aparato matemático subyacente, otro firmado ya por los tres autores, y aún otro en el que Pauli calculaba los términos de la serie de Balmer del hidrógeno con la nueva mecánica.[6] En noviembre, Dirac, ya enterado a través de Fowler del trabajo de Heisenberg, firmó su primer artículo; y un par de meses después, también indicó cómo acometer, con las nuevas herramientas, el problema del hidrógeno.[7] La tercera y última propuesta vino de Zürich, donde en esos momentos vivía y trabajaba Schrödinger. Su primer artículo se publicó en febrero de 1926.[8]

**2.1. Mecánica Matricial, Mecánica Cuántica**

El análisis inicial de Heisenberg no puede ser más claro respecto al nuevo estatuto de las partículas elementales, al menos en el interior del átomo: la teoría sólo tratará con observables. En este sentido, nos obliga a introducir la primera distinción entre el grado de realidad que puede otorgarse a los electrones y a los átomos: respecto a estos últimos, no intuimos siquiera algún vislumbre de duda respecto a su existencia. Por el contrario, queda desterrado cualquier intento de visualizar su interior, de pensarlo en términos de electrones orbitando alrededor del núcleo:[9]

> It is well known that the formal rules which are used in quantum theory for calculating observable quantities such as the energy of the hydrogen atom may be seriously criticized on the grounds that they contain, as basic element, relationships between quantities that are apparently unobservable in principle, e.g., position and period of revolution of the electron. These rules lack an evident physical foundation, unless one still wants to retain the hope that the hitherto foundation unobservable quantities may later come within the realm of experimental determination. This hope might be regarded as justified if the above-mentioned rules were internally consistent and applicable to a clearly defined range of quantum mechanical problems. Experience however shows that only the hydrogen atom and its Stark effect are amenable to treatment by these formal rules of quantum theory.

Heisenberg rubrica pues el fracaso de las teorías ordinarias para dar cuenta de los espectros. Dicho fracaso debe llevarse por delante, en su opinión, la utilización de magnitudes y objetos físicos no observables. El hecho de que el artículo esté dedicado a los espectros explica en buena medida por qué Heisenberg no aborda directamente la cuestión de la constitución última de la materia. Da por hecho que hay átomos, pero da también por hecho que hay electrones, aunque a estos últimos los prive de localización y trayectoria. Ello no quita para que se les pueda atribuir emisión y absorción de radiación:[10]

> … it is necessary to bear in mind that in quantum theory it has not been possible to associate the electron with a point in space, considered as a function of time, by means of observable quantities. However, even in quantum theory it is possible to ascribe an electron the emission of radiation. In order to characterize this radiation we first need the frequencies which appear as functions of two variables.

Semejante planteamiento es también el que encontramos en los artículos subsiguientes con que quedó inicialmente levantada la estructura de la Mecánica Matricial. Y del mismo tono son las contribuciones de Dirac. Su posicionamiento respecto a la cuestión que aquí nos ocupa es muy similar, al menos inicialmente, al de los creadores de la Mecánica Matricial: indiferencia.
En definitiva, la lectura de estos trabajos seminales puede llevar a juzgar nuestra pregunta impertinente, pues en ningún momento parece que los autores se cuestionen el atomismo. Escriben sobre átomos y electrones sin escrúpulos, a pesar de manifestar claramente su adhesión a la ocurrencia heisenbergiana de atender sólo a las magnitudes físicamente observables.
Pasemos a ver ahora cómo Schrödinger sí justifica el tema de nuestra ponencia.

---

[6] BORN & JORDAN (1925), BORN & HEISENBERG & JORDAN (1925), PAULI (1926).
[7] DIRAC (1925, 1926a).
[8] SCHRÖDINGER (1926b).
[9] HEISENBERG (1925). En VAN DER WAERDEN (1967), 261. Cuando esté disponible -como en este caso-, citaremos una traducción inglesa de los artículos escritos originalmente en alemán.
[10] *Ibid.*, 263.

## 2.3 Mecánica Ondulatoria

La irrupción en escena de Schrödinger, en enero de 1926, abrió para muchos un resquicio de esperanza, una posibilidad de reencontrarse con conceptos y explicaciones tradicionales. Y es que como indica su nombre, la Mecánica Ondulatoria alentaba una interpretación de tipo continuista, tanto en lo que se refiere a su negación de la constitución discreta de la materia, como por su cercanía a la mecánica y electrodinámica vigentes hasta entonces. Aunque el propio Schrödinger tuvo que lidiar con varios problemas y cambiar su propia interpretación al menos a lo largo de los primeros años, nunca dejó de defender una revisión de la idea misma de átomo, y cuestionó hasta el final de sus días la misma idea de 'salto cuántico'.

Como hemos dicho, en esta comunicación nos referiremos sólo a sus contribuciones fundacionales.[11] Sin embargo, en el caso de Schrödinger es obligado comentar, aunque sea de pasada, sus antecedentes en relación a los fenómenos cuánticos. Fue, sin duda, el que más tarde se adentró en ese terreno, pero también era, probablemente, uno de los más versados en Física Estadística. De hecho, en diciembre de 1925, y después de cartearse con Einstein al respecto, publicó un interesante y original artículo en el que hacía un tratamiento completo de un gas ideal, pero desde un punto de vista ondulatorio.[12] Esto es, agotó, en términos pre-mecánico cuánticos, la analogía entre materia y radiación a la que Einstein andaba dando vueltas desde hacía más de una década. Lo que hizo Schrödinger fue tratar un gas ideal material como una cavidad con modos propios, obviamente con rasgos que lo diferenciaran de la radiación, como por ejemplo la conservación del número de partículas.

No es este el sitio para analizar en detalle esta contribución de Schrödinger, pero sí merece la pena señalar que ya antes de dar con su propuesta de solución del espectro del hidrógeno era plenamente consciente de que, a cierto nivel, las novedades cuánticas conducían a una difuminación de la frontera que separaba fenómenos ondulatorios y corpusculares. Einstein así lo había manifestado en 1909, y a raíz de ello en 1924 había extendido, en pocas semanas, el trabajo de Bose sobre el cuerpo negro a un gas de partículas, citando además -y dando así a conocer- la tesis de De Broglie. El propio Schrödinger presentó años después su artículo como un trabajo preparatorio, tentativo, previo (pero conectado) a sus artículos sobre Mecánica Ondulatoria.[13]

En cualquier caso, la lectura atenta de esa y otras contribuciones relativamente cercanas tanto en la temática como en el tiempo a la creación de la Mecánica Ondulatoria no parece que permita deducir que ya por entonces Schrödinger se hubiera acogido a una visión continuista de la materia. Eso sí, muestra que estaba abierto a experimentar e investigar visiones alternativas. Tal fue la impresión que le causó la pérdida de independencia que estaban sufriendo las partículas en las estadísticas cuánticas.

Así, en enero de 1926 Schrödinger envió el primer artículo sobre la "Cuantización como un problema de valores propios"[14]. Ni este ni los tres posteriores artículos de la serie estaban dedicados a los problemas estadísticos a que nos acabamos de referir, sino a la forma del espectro del hidrógeno y la estructura atómica en general. El protagonista principal de su presentación es la función de onda, solución de la ecuación que hoy lleva su nombre. Pero su significado físico no fue claro desde un inicio.

En la segunda entrega, Schrödinger planteó la analogía con la que pretendía fundamentar su programa: su mecánica era a la mecánica ordinaria lo que la Óptica Ondulatoria a la Óptica Geométrica. Según eso:[15]

---

[11] SCHRÖDINGER (1926b, 1926c, 1926e, 1926f).
[12] SCHRÖDINGER (1926a).
[13] BACCIAGALUPPI & VALENTINI (2009), 112, nota al pie.
[14] SCHRÖDINGER (1926b).
[15] SCHRÖDINGER (1926c). Versión inglesa en SCHRÖDINGER (1982), 25.

> The study of the motion of image points, which is the object of classical mechanics, is only an approximate treatment, and has, as such, just as much justification as geometrical or "ray" optics has, compared with the true optical process. A macroscopic mechanical process will be portrayed as a wave signal of the kind described above, which can approximately enough be regarded as confined to a point compared with the geometrical structure of the path.

De manera que ya en esta segunda entrega Schrödinger propone entender las representaciones corpusculares como aproximaciones macroscópicas. Un poco más adelante es más explícito, si cabe:[16]

> In *this* sense do I interpret the "phase waves" which, according to de Broglie, accompany the path of the electron; in the sense, therefore, that no special meaning is to be attached to the electronic path itself (at any rate, in the interior of the atom), and still less to the position of the electron on its path.

Pero dicha renuncia nada tiene que ver con la promulgada por Heisenberg, pues Schrödinger cree que es por ese apego a lo corpuscular que se ha decidido abandonar las explicaciones causales, o espacio-temporales:[17]

> … All these assertions systematically contribute to the relinquishing of the ideas of "place of the electron" and "path of the electron". If these are not given up, contradictions remain. This contradiction has been so strongly felt that it has even been doubted whether what goes on in the atom could ever be described within the scheme of space and time. From the philosophical standpoint, I would consider a conclusive decision in this sense as equivalent to a complete surrender. For we cannot really alter our manner of thinking in space and time, and what we cannot comprehend within it we cannot understand at all. There *are* such things - but I do not believe that atomic structure is one of them.

Así que la tabla de salvación que muchos vieron en la mecánica de Schrödinger venía promovida por él mismo. Enseguida veremos que Born primero, y Heisenberg después, contestaron a este posicionamiento de Schrödinger subrayando y enfatizando el carácter acausal e ininteligible (por azaroso) de los procesos elementales para salvaguardar precisamente las partículas elementales. Más en concreto, Heisenberg contestó a una nota que encontramos en el artículo que Schrödinger dedicó a la equivalencia entre las distintas formulaciones de la nueva mecánica:[18]

> My theory was inspired by L. de Broglie (…), and by brief, yet infinitely far-seeing remarks of A. Einstein (…). I did not at all suspect any relation to Heisenberg's theory at the beginning. I naturally knew about this theory, but was discouraged, if not repelled, by what appeared to me as very difficult methods of transcendental algebra, and by the want of perspicuity.

Pero había procesos en los que parecía imposible negar el carácter corpuscular de la materia, como por ejemplo los procesos de colisión. Fue precisamente al tratar de explicarlos con la nueva mecánica cuándo surgieron las primeras discrepancias internas entre los físicos de Göttingen. Schrödinger, por su lado, recurrió a los paquetes de onda: ondas de diferentes frecuencias superpuestas de manera que permitían localizar el objeto estudiado, asemejando así las partículas a singularidades en un sustrato ondulatorio.

Schrödinger planteó y detalló su propuesta con el oscilador harmónico, prácticamente el único que no presenta problemas al respecto. En los meses posteriores, cuando la oposición entre los partidarios de las distintas interpretaciones se enconó, se puso repetidamente de manifiesto que en la mayoría de los casos los paquetes de onda de Schrödinger se dispersan con el tiempo, cosa que, como es sabido, no les pasa a las partículas. De hecho, el propio Schrödinger era consciente, ya en 1926, de que no era posible mantener los paquetes localizados ni en ausencia de campos ni para electrones con números cuánticos muy altos.[19] En Solvay expresó su confianza en solventar ese problema, que de momento sólo se podía trampear con una mezcla de ondas y electrones puntuales.

---

[16] *Ibid.*, 26. Énfasis en el original.
[17] *Ibid.*, 26-27. Énfasis en el original.
[18] SCHRÖDINGER (1926d). Versión inglesa en SCHRÖDINGER (1982), 46.

En relación a la interpretación de la función de onda, en estos primeros artículos Schrödinger se decantó finalmente por interpretarla como densidad de carga y rehuyó la interpretación probabilística. Esto leemos en la cuarta y última entrega, enviada en junio:[20]

> $\Psi\bar{\Psi}$ is a kind of *weight-function* in the system's configuration space. The *wave-mechanical* configuration of the system is a *superposition* of many, strictly speaking of all, point-mechanical configurations kinematically possible. Thus, each point-mechanical configuration contributes to the true wave-mechanical configuration with a certain *weight*, which is given precisely by $\Psi\bar{\Psi}$. If we like paradoxes, we may say that the system exists, as it were, simultaneously in all the positions kinematically imaginable, but not "equally strong" in all.

En la ironía de esta última frase detectamos nuevamente su distanciamiento de las supuestas paradojas a que arrojaban los descubrimientos cuánticos. Según él, provenían sin más de aferrarse a las imágenes corpusculares. Más abajo volveremos sobre la interpretación de Schrödinger, según la cual el cuadrado de la función de onda no es la probabilidad de encontrar una partícula en un lugar con una cierta velocidad, sino que da una especie de superposición de estados que describen lo que le pasa efectivamente al sistema. El sistema no puede pues consistir en partículas, sino en una especie de complejo ondulatorio. Ondas, sí, pero peculiares.

**3. Después de Schrödinger**

La cercanía de los métodos de Schrödinger a los métodos tradicionales y el concomitante éxito de su formulación entre sus colegas vinieron acompañados de cambios importantes en las interpretaciones de los físicos de Göttingen. Dicho viraje se produjo cuando otro tipo de sistemas, distintos de los espectros y de carácter aperiódico, pasaron a ser objeto de estudio.

Todo ello se tradujo en una recuperación de las imaginerías atómicas, en detrimento, eso sí, de las explicaciones causales. Schrödinger no cejó en su empeño de reducir la discretización a la mínima expresión, y se preocupó de indicar cómo dar cuenta de experimentos a la sazón muy conocidos y típicamente corpusculares, como el de Franck-Hertz o el efecto Compton.[21]

**3.1 La interpretación probabilística de Born**

Si alguno de los autores de la Mecánica Matricial reaccionó a la aparición de los trabajos de Schrödinger, ese fue Born. Se dio cuenta de que la formulación del vienés se adaptaba mucho mejor que la matricial al tratamiento de problemas no periódicos.[22] Eso sí, dejó atrás buena parte de los resquemores positivistas, a cambio de matizar o modificar profundamente la vigencia de la causalidad en los procesos elementales. Vamos a comentar brevemente los dos primeros artículos en los que presentó la interpretación probabilística de la función de onda.[23]

El primero es de tipo programático. Según explica él mismo, fue Bohr quién propuso estudiar este aspecto hasta entonces ajeno a la nueva mecánica, con el objetivo de conocer más a fondo los procesos de emisión y absorción, cuya naturaleza era a la sazón completamente ignorada. Como hemos dicho, en este artículo Born admite que la formulación de Schrödinger se adapta mucho mejor a este fenómeno que la de los físicos de Göttingen (entre los que no olvidemos que se cuenta él mismo). Para empezar, hay que eliminar el problema del acoplamiento entre el electrón incidente y el átomo. Born supone que el primero viene y va, asintóticamente, del infinito. Ello permite asociarlo a una onda plana:[24]

---

[19] Véase BACCIAGALUPPI & VALENTINI (2009), 116 y ss.
[20] SCHRÖDINGER (1926f). Versión inglesa en SCHRÖDINGER (1982), 120. Énfasis en el original.
[21] SCHRÖDINGER (1927a, 1927b).
[22] No nos dedicaremos aquí a la intentona con la que, junto a Wiener, trató precisamente de extender la Mecánica Matricial a sistemas aperiódicos: BORN & WIENER (1926). Véase BACCIAGALUPPI & VALENTINI (2009), 93-94.
[23] BORN (1926a, 1926b).

> According to Schrödinger, the atom in its $n$th quantum state is a vibration of a state function of fixed frequency $W_n^0/h$ spread over all of space. In particular, an electron moving in a straight line is such a vibratory phenomenon which corresponds to a plane wave.

¿Cómo interpretar esa onda? Como la densidad de probabilidad de que el electrón salga despedido en una dirección determinada (en una nota al pie añadida en pruebas se decanta ya por que sea el cuadrado de la función de onda el que da esa probabilidad). La descripción causal de por qué el electrón sale en una u otra dirección en uno u otro momento queda, de momento, fuera de la teoría. Si eso llegará a poder determinarse o no es, para Born, una cuestión de otro orden -"filosófico"-, y él prefiere no detenerse en disquisiciones que no vayan acompañadas de artillería matemática. Años más tarde admitió que su familiaridad con los experimentos de colisiones de electrones, que Franck preparaba diariamente en Göttingen en el mismo edificio donde él trabajaba, le habían convencido completamente de la corpuscularidad del electrón.[25]

En la siguiente entrega, mucho más prolija, Born se desmarca de Heisenberg y Schrödinger para presentar su propia interpretación de los procesos atómicos. Aunque prueba su utilidad sólo en los procesos de colisión, cree que es extensible a otros fenómenos. En esta ocasión admite su deuda con la idea de Einstein, inspirada a su vez en la de De Broglie, del "campo fantasma", que sería la onda que guia al fotón. Él aplica esa idea al electrón:[26]

> I should therefore like to investigate experimentally the following idea: the guiding field, represented by a scalar function $\psi$ of the coordinates of all the particles involved and the time, propagates in accordance with Schrödinger's differential equation. Momentum and energy, however, are transferred in the same way as if corpuscles (electrons) actually moved. The paths of these corpuscles are determined only to the extent that the laws of energy and momentum restrict them; otherwise, only a probability for a certain path is found, determined by the values of the $\psi$ function. This could be summarized, somewhat paradoxically, perhaps as follows: the motion of the particles follows laws of probability, but that probability itself propagates in harmony with the causal law.

Queda claro que Born ha recaído en el empleo de visualizaciones, y no hallamos tampoco rastro de duda de la corpuscularidad de los electrones. De hecho, celebra esa recaída, a pesar de que conlleve una enmienda a la causalidad:[27]

> On that basis of the above discussions, I should like to put forward the opinion that quantum mechanics permits not only the formulation and solution of the problem of stationary states, but also that of transition processes. In these circumstances, Schrödinger's version appears to do justice to the facts in by far the easiest manner; moreover, it permits the retention of the conventional ideas of space and time in which events take place in a completely normal manner. On the other hand, the proposed theory does not correspond to the requirement of the causal determinacy of the individual event.

Born abundó en las mismas ideas en un artículo posterior que publicó en la primavera de 1927, titulado "Quantenmechanik und Statistik".[28] Ahí descarta la propuesta de Schrödinger de asociar paquetes de onda a las partículas. Pero aunque los corpúsculos carezcan de individualidad (Born se refiere ya al concepto de 'indistinguibilidad'), podemos seguir imaginándonoslos:[29]

---

[24] BORN (1926a). Versión inglesa en WHEELER & ZUREK (1983), 53.
[25] JAMMER (1974), 39.
[26] BORN (1926b). Versión inglesa en LUDWIG (1968), 207.
[27] *Ibid.*, 224.
[28] BORN (1927).
[29] *Ibid.* En BORN (1963), 304.

> Die Materieselbst kann nach wie vor unter dem Bilden beweglicher (punktförmiger) Teilchen (Elektronen, Protonen) vorgestellt werden; nur sind diese Korpuskeln in vielen Fällen gar nicht als Individuen zu identifizieren, z. B. dann, wenn sie zu einem Atomverband zusammentreten.

Pero no ahonda en las posibles contradicciones, dejando una vez más la cuestión en manos de aquellos que quieran filosofar.

Unos meses después, Pauli, en un trabajo en el que aplicaba la estadística de Fermi-Dirac al estudio del paramagnetismo, extendió la interpretación probabilística a los gases ideales degenerados.[30] De nuevo, el hecho de que éste recurriera a la función de onda de Schrödinger, no implicaba en absoluto que compartiera su interpretación continuista. De hecho, tanto él como Heisenberg se encargaron de mostrar que los resultados de Born podían deducirse también en el marco de la Mecánica Matricial, sin recurrir a la función de onda.[31]

La interpretación probabilística empezó a establecerse como uno de los resultados más característicos e impactantes de la nueva mecánica. La función de onda evolucionaba causalmente, según la ecuación de Schrödinger. Pero su colapso no seguía ley alguna que no fuera estadística. Por otro lado, el hecho de que no estuviera definida en el espacio tridimensional, como una onda ordinaria, seguía impidiendo su visualización. Dicha polidimensionalidad contenía, trasladada a las tres dimensiones, el acoplamiento o resonancia entre sistemas; lo que más adelante se denominará entrelazamiento.

A Schrödinger, como era de esperar, este uso de la función de onda no le convenció:[32]

> From an offprint of Born's last work in the Zeitsch. f. Phys. I know more or less how he thinks of things: the waves must be strictly causally determined through field laws, the wavefunctions on the other hand have only the meaning of probabilities for the actual motions of light- or material particles. I believe that Born thereby overlooks that… it would depend on the taste of the observer which he now wishes to regard as real, the particle or the guiding field.

### 3.2 La trayectoria de Heisenberg

Heisenberg también desarrolló su propia interpretación de la teoría una vez Schrödinger hubo puesto en danza su Mecánica Ondulatoria. Para ver ese cambio con un poco de detalle, y cómo ese proceso hizo que tomara partido contra la visión continuista, comentaremos tres trabajos, de 1926 y 1927, siendo el último aquel en que aparece por primera vez el principio de incertidumbre.[33]

En verano de 1926 Heisenberg se dedicó al problema de varios cuerpos, en parte para aplicarlo al problema del Helio. En su primer trabajo al respecto presenta lo que viene a ser la versión mecánico cuántica de la indistinguibilidad que ya Bose y Einstein habían introducido más de un año antes, pero con el formalismo de la Mecánica Estadística clásica, en el espacio fásico.[34] Heisenberg analiza en este artículo la "resonancia", fenómeno que en los años 30 Schrödinger bautizaría como "entrelazamiento" ("Verschränkung").

Heisenberg ve como un punto débil de la Mecánica Ondulatoria precisamente que necesite recurrir más a la visualización que la Matricial.[35] Pero lo que más le interesa destacar es que la vía de Schrödinger no debe considerarse la continuación de la de De Broglie, por el mero hecho de que las ondas tridimensionales propuestas por éste último no se corresponden con las ondas de más dimensiones de Schrödinger. Una descripción con ondas tridimensionales recuperaría una descripción tradicional espacio-temporal, lo que Heisenberg, a estas alturas, considera ya una rémora del pasado. Así, este artículo de Heisenberg puede leerse como una reivindicación de las ideas de De Broglie. O sea, un pronunciamiento de que es la Mecánica Matricial, y no la Ondulatoria, la legítima heredera de la vía correcta de formalización y generalización de las ideas del físico francés. Veremos que Jordan hará lo mismo respecto a la hipótesis cuántica de los Planck, Einstein y Bohr.

---
[30] PAULI (1927), 83, nota 1.
[31] Véase BACCIAGALUPPI & VALENTINI (2009), 98 y ss.
[32] Carta de Schrödinger a Wien, 25 de agosto, 1926. Citada en *ibid.*, 118.
[33] HEISENBERG (1926a, 1926b, 1927).
[34] HEISENBERG (1926a).
[35] Él no la llamaba Mecánica Matricial. Al parecer, no le gustaba tal denominación. Véase BACCIAGALUPPI & VALENTINI (2009), 81, nota al pie.

Sin embargo, Heisenberg apela a la corpuscularidad para privilegiar la Mecánica Matricial:[36]

> In Abetracht dieser Analogie scheint es mir eine der wichtigsten Seiten der Quantenmechanik, daβ die auf der Korpuskularvorstellung der Materie basiert ist; freilich handelt es sich dabei nicht um eine Beschreibung der Bewegungen von Korpuskeln in unseren gewöhnlichen Raum-Zeitbegriffen. Dies konnte man auch kaum erwarten; denn selbst wenn sich die Korpuskeln als Singularitäten der metrischen Struktur des Raumes herausstellen sollten, wie es der Wunsch der Kontinumstheorien ist, so wäre dies wohl keine Beschreibung in unseren gewöhnlichen Raum-Zeitbegriffen - es sei denn, man rechnet einen Raum, dessen Maβbestimmung von der Euklidischen wesentlich abweicht, zu den „gewöhnlichen" Räumen.

Es decir, que mantiene los corpúsculos a pesar de negarles visualización. Ahora bien, del mismo modo que las ondas tridimensionales no pueden dar cuenta de la información contenida en la función de onda, tampoco pueden los corpúsculos individuales. Hay que suponerles cierta interrelación, acoplamiento, resonancia:[37]

> Wollte man sich von der Bewegung der Elektronen im Atom ein der quantenmechanischen Lösung einigermaβen entsprechendes anschauliches Bild machen, so müßte man sich hier etwa vorstellen, daβ die beiden Elektronen periodisch in kontinuierlicher Weise die Plätze tauschen, in Analogie zu den Energieschwebungen beim oben erwähnten Oszillatorbeispiel, wobei die Periode dieser Schwebung eben durch den Abstand des Orthoheliumterms vom entsprechenden Parheliumterm gegeben ist.

A estas alturas, podemos decir que si bien Heisenberg quería mantener distancias con la interpretación continuista de Schrödinger, para lo cual resaltó las discontinuidades que respetaba y en las que se fundaba la Mecánica Matricial, vemos que ello no significa tampoco que defendiera a ultranza, ni mucho menos, una cosmovisión atomista. Así le manifestaba a Pauli la repulsa que le provocaba la idea de Born:[38]

> Übrigens was sagen Sie zu Borns letzter Note in der Zeitschrift?? Ein Satz erinnerte mich lebhaft an ein Kapitel aus dem christl[ichen] Galubensbekenntnis: „Ein Elektron *ist* eine ebene Welle..." Sehr schön ist es auch, wenn mann über den tieferen Sinn der Fuβnote S. 865 nachdenkt. Aber ich will Ihnen im Lästern keine Konkurrenz machen.

La nota al pie a la que se refiere es aquella en la que aparece por primera vez -que nosotros sepamos- la propuesta de interpretar el cuadrado de la función como una densidad de probabilidad.

Sus reticencias a mantener una visión corpuscular simple son evidentes en la transcripción de una charla que pronunció en la reunión anual de la *Naturforscherversammlung* en Düsseldorf, el 23 de septiembre de ese mismo año.[39] Este documento representa, en nuestra opinión, el último hito en la trayectoria de Heisenberg antes de su viraje ya más claramente orientado a visualizar los procesos intratómicos.

De hecho, en esta interesante ponencia, Heisenberg considera la relación de la nueva mecánica con la cuestión del atomismo. Presenta tres restricciones a la realidad de los

---

corpúsculos, tres argumentos contra la afirmación de que la materia está hecha de primordios. Son los siguientes:

I. Los intentos de explicación de los fenómenos microscópicos han fracasado por consistir en la confección de imágenes. Esta sería la primera limitación: la *imposibilidad de visualizar los átomos*.
II. La Mecánica Matricial define magnitudes análogas a las clásicas, como la posición o el momento. Pero si bien puede asociarse energía a los estados estacionarios, por ejemplo, *las magnitudes observables asociadas a un corpúsculo tampoco son directamente visualizables*, concebibles.
III. La última restricción es la que ya establecieron Bose y Einstein y que Heisenberg mismo se encargó de extender a la nueva mecánica: la *indistinguibilidad de las partículas*. Para dar cuenta de un sistema de varios cuerpos, hay que suponer que éstos están intercambiando constantemente sus posiciones. A nivel mecánico cuántico no hay más que suponer que sólo son válidas un tipo de soluciones de la ecuación de Schrödinger (las simetrizadas), que tratan el sistema en conjunto, como un todo. Los corpúsculos no poseen un grado de realidad que les asegure la identidad.

Hasta aquí su afrenta al atomismo. ¿Significa esto que Heisenberg se estaba repensando su repudio a las ondas de materia? En absoluto. Heisenberg nunca puso las ondas y las partículas en pie de igualdad, y señaló la asimetría entre ambos tipos de explicación: si bien hay algunos experimentos que admiten explicaciones corpusculares u ondulatorias, determinados fenómenos sólo admiten explicaciones corpusculares, y nunca ondulatorias.

Schrödinger estuvo en Copenhague en el mes de octubre[40], una vez ya se habían publicado las 4 entregas de su mecánica. Pero antes, en julio, había dado charlas en Berlín y Múnich.[41] Seguramente a esta última visita se refería Heisenberg en una carta a Pauli que ya hemos citado antes. En ella, su oposición a la versión ondulatoria es manifiesta:[42]

> So nett Schr[ödinger] personlich ist, so merkwürdig find'ich seine Physik: man kommt sich, wenn man sie hört, um 26 Jahre jünger vor. Schr[ödinger] wirft ja alles „quantentheoretische": nämlich lichtelektrischen Effekt, Francksche Stöβe, Stern-Gerlacheffekt usw. über Bord; dann ist es nicht schwer, eine Theorie zu machen. Aber sie stimmt eben nicht mit der Erfahrung.

Tal era la disposición de Heisenberg cuando Schrödinger atendió a la invitación de Bohr. En Copenhague tuvieron lugar encendidas discusiones entre Schrödinger, Bohr y Heisenberg, que probablemente contribuyeron a perfilar las posturas de los participantes. En lo que respecta a Heisenberg, en el siguiente artículo que comentaremos, escrito en marzo del año siguiente, presenta una postura aún más beligerante con el continuismo de Schrödinger. El contraste con el que acabamos de ver es muy llamativo. Podemos apreciar en una carta previa a Pauli, de finales de diciembre, cómo el posicionamiento de Heisenberg es más bien negativo (se niega a admitir el continuo), en consonancia con su análisis del atomismo en el mundo cuántico:[43]

> Daβ die Welt kontinuierlich sei, halte ich mehr denn ja für gänzlich indiskutabel. Aber sobald sie diskontinuierlich ist, sind in allen unseren Worten, die wir zur Beschreibung eines Faktums verwenden, zu viele c- Zahlen. Was das Wort „Welle" oder „Korpuskel" bedeutet, weiβ man nicht mehr.

Nos referiremos ahora al artículo donde Heisenberg presenta el principio de incertidumbre.[44] Lo que encontramos en él es un intento de interpretar los resultados de la Mecánica Matricial a partir de los márgenes de incertidumbre en las medidas físicas. Así, en cierto modo, se puede volver a hablar de la trayectoria del electrón. Obviamente no en el sentido hasta entonces usual, pero varios de los experimentos mentales que Heisenberg presenta exigen una visualización que no es compatible con su propuesta rompedora de 1925. En un punto del artículo, Heisenberg considera incluso las dimensiones del tamaño del electrón; eso sí, libre.

---

[40] Consta que el 4 de octubre dio una charla en Copenhage. Véase PAULI (1979), 339.
[41] Véase JAMMER (1974), 31.
[42] Carta de Heisenberg a Pauli, 28 de julio, 1926. En PAULI (1979), 337-338.
[43] Carta de Pauli a Heisenberg, 23 de noviembre, 1926. En *ibid.*, 359. "c-Zahlen" alude a los números clásicos, magnitudes de la física clásica.
[44] HEISENBERG (1927).

También en contra de una teoría eminentemente estadística, Heisenberg asocia el problema de la causalidad con el problema de la medición: no puede predecirse con exactitud qué ocurrirá porque no puede conocerse el presente con exactitud. Si detrás de los observables hay un mundo determinista o no, forma parte de la especulación. A partir de los experimentos, hay que afirmar que la conexión causal no rige en los procesos atómicos. Y eso es lo que pone de manifiesto el flamante principio de incertidumbre.

La creciente animadversión a la Mecánica Ondulatoria en este punto vuelve a ser patente. Veamos una nota que pertenece a este artículo y que parece una respuesta directa al texto de Schrödinger de nuestra nota 18:[45]

> Schrödinger describes quantum mechanics as a formal theory of frightening, indeed repulsive, abstractness and lack of visualizability. Certainly one cannot overestimate the value of the mathematical (and to that extent physical) mastery of the quantum-mechanical laws that Schrödinger's theory has made possible. However, as regards questions of physical interpretation and principle, the popular view of wave mechanics, as I see it, has actually deflected us from exactly those roads which were pointed out by the papers of Einstein and de Broglie on the one hand and by the papers of Bohr and by quantum mechanics on the other hand.

En este apunte vemos una vez más cómo Heisenberg quiere desmentir la visión de que la Mecánica Ondulatoria es la heredera de las ideas de Einstein y De Broglie. Es precisamente su propuesta la que no se desvía de la buena dirección. Y eso tiene más sentido justamente en este artículo, en el que Heisenberg vuelve a hablar de corpúsculos, colisiones y trayectorias. Además de la crítica al espacio de configuraciones y a sus ondas invisualizables, en este nuevo ataque Heisenberg desarbola para siempre el intento de Schrödinger de representar partículas mediante paquetes de onda. Muestra que su creciente dispersión hace que la partícula se deslocalice paulatinamente sin necesidad de influencia externa alguna. El único sistema en que eso no ocurre es el oscilador harmónico. Fuera de eso, el dispositivo ideado por el físico vienés para hacer compatible su visión continuista con los experimentos de dispersión o emisión de partículas no es operativo.

Antes de acabar, merece la pena mencionar una cuestión que otros historiadores ya han comentado antes que nosotros.[46] Este artículo lleva un apéndice añadido en pruebas de imprenta, y que no es sino una consecuencia de las discrepancias entre Heisenberg y Bohr. Éste le reprochó a aquél su dejadez para con los aspectos ondulatorios de alguno de los experimentos descritos. A pesar de que las ideas de Bohr no entren en nuestras consideraciones, no podemos dejar de comentar, aunque sea de pasada, que la complementariedad que ya estaba barruntando el físico danés tras las discusiones con Schrödinger, seguramente le llevó a ver en el artículo de Heisenberg un planteamiento demasiado arrimado a la corpuscularidad. Demasiado anti-ondulatorio y lejano a la dualidad que propugnaría en pocos meses y pasaría a convertirse en la ortodoxia. Dichas discrepancias tenían pues como eje principal el no privilegiar ninguno de los dos aspectos de la dualidad. Esto le escribía Heisenberg a Pauli en mayo de 1927:[47]

> Bohr will eine allgemeine Arbeit über den „begrifflichen Aufbau" d[er] Qu[anten]th[eorie] schreiben unter dem Gesichtspunkt: „Es gibt Wellen und Korpuskeln", -wenn man damit gleich anfängt, kann natürlich auch alles widerspruchsfrei machen... Trotzdem bin ich natürlich nach wie vor der Ansicht, daβ die Diskontinuitäten das einzig interessante an der Qu[anten]th[eorie] sind und daβ man sie nie genug betonen kann...

---

[45] *Ibid.* Versión inglesa en WHEELER & ZUREK (1983), 82, nota al pie.
[46] ROSENFELD (1971).
[47] Carta de Heisenberg a Pauli, 16 de mayo, 1927. En PAULI (1979), 394-395.

## 3.3 La indistinguibilidad de Dirac

Vimos más arriba que los escritos de Dirac se caracterizan por la preeminencia de las relaciones meramente matemáticas y el descuido de su interpretación física. Aunque dichos planteamientos suelen tildarse de asépticos, neutros, eso no siempre está justificado. Ya comentamos cómo Dirac se refería, en esos artículos de 1925, a átomos y electrones sin problematizarlos. En 1926, publicó otro artículo crucial para el desarrollo de la Mecánica Cuántica en el que se hace más notorio que para Dirac romper con la Mecánica Clásica no tenía nada que ver con romper con el atomismo.[48]

En verano envió a los *Proceedings* de la *Royal Society* un trabajo en el que consideraba, entre otros, el problema del gas ideal y también del problema de las colisiones. Así, casi al mismo tiempo que Heisenberg, enfocaba su atención al sistema de varios cuerpos, el más sencillo de los cuáles es el del gas de partículas sin interacción. A diferencia del físico de Göttingen, Dirac sí identificó correctamente las funciones simétricas con la estadística de Bose-Einstein, y las antisimétricas con las de Fermi-Dirac. Es interesante señalar que probablemente Dirac fue el primero en utilizar la terminología "estados físicamente indistinguibles". Aunque en este artículo no se refiere directamente a partículas indistinguibles, Dirac siempre explica y ejemplifica este concepto planteando intercambios de electrones. Por ejemplo: "The question arises whether the two states (*mn*) and (*nm*), which are physically indistinguishable as they differ only by the interchange of the two electrons, are to be counted as two different states or as only one state…?".[49] Recordemos que esa era una de las limitaciones de la realidad atómica que Heisenberg presentó en setiembre ante sus colegas alemanes. Pero Dirac da cuenta del principio de exclusión de Pauli en términos netamente corpusculares:[50]

> An antisymmetrical eigenfunction vanishes identically when two of the electrons are in the same orbit. This means that in the solution of the problem with antisymmetrical eigenfunctions there can be no stationary states with two or more electrons in the same orbit, which is just Pauli's exclusion principle. The solution with symmetrical eigenfunctions, on the other hand, allows any number of electrons to be in the same orbit, so that this solution cannot be the correct one for the problem of electrons in an atom.

Una vez más, pero ahora de una manera un poco más flagrante, Dirac pasa por encima de un tema conceptualmente complicado, en el que el propio Heisenberg había reconocido dificultades, sin hacer mención expresa de ello. Recordemos, además, que el tema no era nuevo, pues Heisenberg y Dirac lo que hicieron fue trasladar a la nueva mecánica los resultados ya conocidos desde la publicación de los trabajos de Bose, Einstein y Fermi.

De modo que, a pesar de prescindir de imágenes y visualizaciones, y de que el intercambio de electrones no se traduzca en estados físicamente distintos, Dirac, a diferencia de Heisenberg, no da muestras de dudar de su realidad.

## 3.4 La dualidad de Jordan

Nos referiremos, para concluir, a otro de los físicos de Göttingen, seguramente el menos estudiado: Jordan. Éste hizo mucho por perfilar la idea de dualidad. Dualidad que en su caso no conllevaba negación alguna de la realidad de los átomos y electrones. Así, aunque por momentos parece que convierta a las ondas de probabilidad en el objeto físico mecánico-cuántico, no deja de hablar de los electrones y su trayectoria. En cierto modo, Jordan representa un caso eximio de adhesión a la interpretación de Born.

En la que fuera su *Habilitationsvortrag* en Göttingen, a finales de 1926, hizo mucho hincapié en la acausalidad de los procesos elementales, en la prioridad de las probabilidades en las ecuaciones.[51] Si bien admite que dichas probabilidades son imposibles de entender, dada su definición en el espacio de configuraciones y su irreductibilidad a procesos individuales, Jordan no deja de referirse a las expectativas de encontrar electrones en tal o cual sitio. Propone la interpretación de colectividades:[52]

---
[48] DIRAC (1926b).
[49] *Ibid.* En DIRAC (1995), 185.
[50] *Ibid.*, 187-188.

> In recent times important advances have been made in the discovery of these laws. One can now, for example, compute (in principle) the spectrum connected with the motion of electrons in an atom with the same assurance as, on classical dynamics, one could calculate the motions of the planets. But in spite of the analogy between the calculations, there is an important difference in the interpretation of their results. The classical calculation gives us information about our specific system of planets. The quantum theoretical calculation does not, in general, tell us anything about a single atom, but only about the mean properties of an assembly of similar atoms.

Jordan menciona incluso las trayectorias de las cámaras de niebla: "… we can, in fact, largely because of the work of C.T.R. Wilson, actually observe the fate of a single α-particle, follow its trajectory, and determine the moment when the trajectory ends in a quantum jump".[53] Pero insiste, en los últimos párrafos, en que el tipo de probabilidades que intervienen en la teoría cuántica no se dejan entender, al no poderse reducir a probabilidades de sucesos elementales, estadísticamente independientes.

Así, Jordan aboga por una interpretación corpuscular basada en la interpretación probabilística de la función de onda, descartando por completo los paquetes de onda de Schrödinger. En relación al problema del gas ideal, del que era buen conocedor -antes incluso del advenimiento de la Mecánica Cuántica había trabajado en ello- se refiere a los trabajos de Heisenberg y Dirac sobre la simetrización de la función de onda, a que antes hemos aludido, como a tratamientos corpusculares: "Die Quantenmechanik macht es aber möglich, wie Heisenberg und Dirac gezeigt haben, auch vom rein korpuskulartheoretischen Standpunkt aus die Einsteinsche Gastheorie zu verstehen".[54] La indistinguibilidad de las partículas ("gleicher Teilchen") no sólo no le hace poner en cuestión el atomismo, sino que lo confirma. También el hecho de que un electrón libre venga representado por una onda plana:[55]

> Nehmen wir dieses beispielsweise zunächst als einen kräftefrei bewegten Massenpunkt an, so ist dieser Massenpunkt, falls wir ihm eine scharf definierte Energie, oder, was auf dasselbe hinauskommt, einen scharf definierten Wert des translatorischen Impulses zuschrieben, nach De Broglie darzustellen durch eine rein periodische ebene Welle. Diese Welle hat also überall gleiche Intesitäten in dem ganzen eindimensionalen Raume, in dem sich der Massenpunkt bewegt. Dies legt folgende Vorstellung nahe: Wenn der quantenmechanische Massenpunkt eine scharf definierten Wert des Impulses besitz, so besteht gleichzeitig eine gewisse Wahrscheinlichkeit dafür, daβ sein Ort irgendwo in dem ihm zur Verfügung stehenden Raume ist, und alle möglichen Orte sind durchaus gleichwahrscheinlichkeit.

En Jordan hallamos pues muchas de las características de la que luego será la interpretación ortodoxa. También en el sentido de contradictoria, pues no pretende haber traicionado los principios positivistas de los primeros artículos escritos junto a Born y Heisenberg:[56]

> Alle Deutungen physikalischer Gesetze durch „anschauliche" Bilder geben nichts anderes, als Erläuterungen dieser Gesetze durch Analogien aus dem Bereiche der sichtbaren, greifbaren Dinge. Aber die Gesetze der Mikrophysik sind von so besonderer uns eigentümlicher Art, daβ es nicht möglich ist, für sie vollständige Analogien in der Makrophysik wiederzufinden.

---

[51] JORDAN (1927b)
[52] *Ibid.*, 567.
[53] *Ibid.*, 569.
[54] JORDAN (1927c), 638.
[55] *Ibid.*, 647.
[56] *Ibid.*, 648.

Pero ya hemos visto que ni el propio Jordan lleva hasta el final las consecuencias de su planteamiento. Por otro lado, en su propia contribución al tratamiento de las colisiones con el formalismo matricial, Jordan había reclamado ser uno de los continuadores genuinos de la física cuántica de los Planck, Einstein y Bohr, de los cuáles Schrödinger, en su celo continuista, se habría separado "radicalmente":[57]

> Bekanntlich hat Schrödinger, ausgehend von den mathematischen Beziehungen, durch deren Aufdeckung er die Quantenmechanik bereichert hat, Vorstellungen entwickelt, die sich zu den von Planck, Einstein, Bohr entwickelten Grundannahmen der Quantentheorie in radikalem Gegensatz befinden. Die folgenden Ausführungen gründen sich jedoch durchaus auf die alten Vorstellungen (stationäre Zustände; Übergänge), welche sie genauer zu analysieren und zu stützen geeignet scheinen.

## 4. El quinto Congreso Solvay

En la celebración del Congreso Solvay daremos fin a nuestra comunicación, principalmente porque ahí es donde suele situarse la primera presentación pública más o menos elaborada y acabada de la complementariedad de Bohr. La interpretación de Copenhague, pues, empezó a fraguarse por esa época. En cualquier caso, hasta aquí es donde queríamos asomarnos para ver hasta qué punto el atomismo había calado y hasta qué punto se puso en duda entre los investigadores del mundo cuántico.

De los autores a que nos hemos referido, sólo Born, Heisenberg y Schrödinger presentaron ponencia en Bruselas. Pero al congreso acudieron también Einstein, Pauli, De Broglie, Lorentz y Bohr, entre otros.[58] Vamos a limitarnos aquí a dar unas pinceladas para poner de manifiesto la magnitud de la disparidad de interpretaciones existentes, ya en 1927.

En la comunicación que presentaron conjuntamente Heisenberg y Born no apreciamos novedades (siempre en relación a la pregunta que nos hemos formulado), aunque las posturas sí parecen un poco más asentadas. La declaración de principios aparece casi al comienzo:[59]

> Deux espèces de discontinuités sont caractéristiques pour la physique de l'atome: l'existence de corpuscules (électrons, quanta de lumière) d'une part, l'existence d'états stationnaires séparés (valeurs déterminées de l'énergie, valeurs de l'impulsion, etc.) d'autre part. Les deux espèces de discontinuités ne peuvent être introduites dans la théorie classique que par des hypothèses auxiliaires fort artificielles.

El hecho de que Heisenberg y Born pongan al mismo nivel la corpuscularidad y la cuantización de los estados estacionarios sí sugiere que no estaban sino respondiendo, una vez más, a la cosmovisión de Schrödinger. Pero todo ello no les impide promover la dualidad: "A la nature dualiste de la lumière -ondes, quanta de lumière- correspond la nature dualiste analogue des particules matérielles. Celles-ci aussi se comportent à un certain point de vue comme des ondes".[60]

Por su parte, Schrödinger tampoco dice nada que no hubiera dicho ya. Insiste en su interpretación:[61]

> Le système classique de points matériels n'existe pas réellement, mais il existe quelque chose qui remplit continûment tout l'espace et dont on obtiendrait une «photographie instantanée» si, en laissant ouvert l'obturateur de la chambre noire, on faisait passer le système classique par *toutes* ses configurations, en laissant l'image dans l'espace $q$ séjourner dans chaque élément de volume $d\tau$ pendant un temps qui est proportionnel à la valeur *instantanée* de $\psi\psi$*. (...) Autrement dit: le système réel est une image composite du système classique dans tous ses états possibles, obtenue en employant $\psi\psi$* comme «fonction de poids».

---

[57] JORDAN (1927a), 661, nota al pie.
[58] Las actas pueden consultarse en SOLVAY (1928).
[59] SOLVAY (1928), 144.
[60] *Ibid.*, 164.
[61] *Ibid.*, 191-192. Énfasis en el original.

Pero al parecer no se le entendió demasiado, o eso le debió de parecer a él porque en la discusión posterior vuelve a explicar su visión con motivo de que no parece haber hecho fortuna. En definitiva, Schrödinger se desmarca una vez más de la interpretación probabilista de Born: la evolución de la función de onda no da probabilidades sino el comportamiento del objeto físico real, de aquello que está ocurriendo. Compara su propuesta con la de De Broglie, quien en Bruselas defendió su onda piloto, una onda tridimensional que guiaría a las partículas en su recorrido. El físico francés presentó una especie de propuesta mixta, más bien corpuscular, pero cuyo carácter ondulatorio no era ni mucho menos secundario.

Pero si algo se recuerda de esta reunión es el inicio (o uno de los primeros hitos) del debate entre Einstein y Bohr. Así, a las discrepancias entre los Heisenberg y Born y Schrödinger, la propuesta de De Broglie, y la complementariedad de Bohr, hay que añadir la crítica de Einstein, posteriormente asociada a su postura realista, ligeramente distinta de la de Schrödinger. Lorentz, presidente del Congreso y patriarca de la Física al menos hasta cinco o diez años antes, también manifestó su apego a la visión más clásica del electrón:[62]

> Pour moi, un électron est un corpuscule qui, à un instant donné, se trouve en un point déterminé de l'espace, et si j'ai eu l'idée qu'à un moment suivant ce corpuscule se trouve ailleurs, je dois songer à sa trajectoire, qui est une ligne dans l'espace. Et si cet électron rencontre un atome et pénètre, et qu'après plusieurs aventures il quitte cet atome, je me forge une théorie dans laquelle cet électron conserve son individualité; c'est-à-dire que j'imagine une ligne suivant laquelle cet électron passe à travers cet atome. Il se peut, évidemment, que cette théorie soit bien difficile à développer, mais a priori cela ne me paraît pas impossible.
>
> Je me figure que, dans la nouvelle théorie, on a encore de ces électrons. Il est possible, évidemment, que dans la nouvelle théorie bien développée, il soit nécessaire de supposer que ces électrons subissent des transformations. Je veux bien admettre que l'électron se fond en un nuage. Mais alors je chercherai à quelle occasion cette transformation se produit. Si l'on voulait m'interdire une pareille recherché en invoquant un principe, cela me gênerait beaucoup. Il me semble qu'on peut toujours espérer qu'on fera plus tard ce que ne pouvons pas encore faire en ce moment…

A pesar, pues, de su creencia en la corpuscularidad del electrón, identificado apenas 30 años atrás, Lorentz admite que pueda sufrir transformaciones. Pero exige poder hacerse imágenes, un cómo y un cuándo. Al contrario que Bohr, quien renunció a la posibilidad de usar exclusivamente uno de los dos tipos de imágenes: ondulatorio y corpuscular. Con todas las contradicciones que ello implica, reconocidas por él mismo. En relación a su constante empleo del concepto de 'partícula' en su ponencia, escribió:[63]

> In the above examples we have repeatedly mentioned the velocity of a particle. The purpose, however, was only to obtain a connection with our ordinary space-time description convenient in this case, since, strictly speaking, an unambiguous definition of velocity is excluded by the quantum postulate. This is particularly to be remembered when comparing the results of successive observations.

O más crudamente, en el interior del átomo:[64]

> In fact, the consistent application of the concept of stationary states excludes, as we shall see, any specification regarding the behavior of the separate particles in the atom. In problems, where a description of this behavior is essential, we are bound to use the general solution of the wave equation which is obtained by superposition of characteristic solutions.

---

[62] *Ibid.*, 248-249.
[63] COMO (1928), 574. Citamos en este caso la versión inglesa de la ponencia que Bohr envió a publicar a las actas del Congreso Solvay, y que es la misma que envió, en inglés, a las actas del Congreso Internacional de Física celebrado en Como (Italia) un mes antes.
[64] *Ibid.*, 581.

Bohr llega incluso a descartar la idea misma de salto cuántico o estado estacionario, que para él no es más que una forma de hacerse una idea de algo de lo que uno no debería hacerse una idea. Y lo mismo ocurre con la noción de partícula:[65]

> Summarising, it might be said that the concepts of stationary states and individual transition processes within their proper field of application possess just as much or as little «reality» as the very idea of individual particles. In both cases we are concerned with a demand of causality complementary to the space-time description, the adequate application of which is limited only by the restricted possibilities of definition and observation.

Bohr establece una distinción entre partículas que interaccionan y partículas libres. Sólo en este último caso tiene sentido imaginárselas, pues cuando interaccionan entre ellas, como en el interior del átomo, su acoplamiento (resonancia, entrelazamiento, etc.) no permite pensarlas aisladamente. En las colisiones, esto sólo se aplicaría al momento mismo del choque, que aún y siendo el momento decisivo del proceso, queda completamente fuera del conocimiento del observador (experimental o teórico).

Pero -como insinúa el propio Bohr- esta cuestión deja al descubierto una consecuencia con la que debe acarrear la teoría cuántica. Si la observación modifica el sistema estudiado, y la teoría sólo puede dar cuenta de las medidas, de los observables, el mismo concepto de partícula libre en Mecánica Cuántica se queda en tierra de nadie. A pesar de ello -y siempre según Bohr-, es una abstracción "indispensable for a description of experience in conexión with our ordinary space-time view"[66].

## 5. Comentarios finales

Esta última mirada a la ponencia de Bohr nos devuelve a la pregunta que nos hacíamos al principio: ¿se llegó a poner en duda el atomismo? El físico danés creía que el grado de realidad del concepto de partícula se había visto seriamente mermado por las últimas investigaciones. No hemos encontrado ninguna declaración tan contundente en los escritos de Heisenberg, Born, Jordan o Dirac. Sí de Schrödinger. Este último cuestionó abiertamente la constitución discreta de la materia, y así se mantuvo hasta el final de sus días, a pesar de no dar con una formulación satisfactoria de la nueva teoría.

Estamos parcialmente de acuerdo con la tesis de Mara Beller, según la cual el rápido posicionamiento de Schrödinger tuvo como consecuencia que los físicos de Göttingen se decantaran hacia una ontología de tipo más corpuscular, llegando a contradecir palmariamente sus propios propósitos de pocos meses antes.[67] Pero hay que añadir que, como hemos mostrado, es necesario considerar que el tipo de problemas que trataron en la segunda tanda de 1926 y 1927, y ya por separado, eran harto distintos del simple átomo de hidrógeno de los artículos fundacionales. Así, no nos consta que su apuesta por los observables nunca hubiera afectado realmente a la partícula libre.

Visto lo visto, puede ser útil establecer un orden respecto al grado de reflexión sobre la nueva teoría: Schrödinger-Bohr-Heisenberg-Jordan-Born-Dirac. En último lugar tendríamos a Dirac, el menos dado a interpretar las ecuaciones. En el otro extremo, Schrödinger, para quién un buen físico debería tratar de entender la realidad causalmente y mediante imágenes espacio-temporales, y quien no concebía una Ciencia sólo basada en métodos matemáticos. Se echa de ver que el cuestionamiento del atomismo se dio más en aquellos que más vueltas le dieron a los fundamentos conceptuales de la nueva mecánica. Según esto, nuestra propuesta sería la siguiente: sin duda, la Mecánica Cuántica problematiza el atomismo, pero la mayoría de sus fundadores ni siquiera se apercibieron de ello.

---

[65] *Ibid.*, 587.
[66] *Ibid.*, 568.
[67] BELLER (1999).

Después de esta etapa inicial, hubo pocas voces que se aunaran a la de Schrödinger. No deja de ser sintomático el desigual trato que han recibido en la divulgación y en la filosofía de la ciencia las discrepancias de Bohr con Einstein y las de los físicos de Göttingen con Schrödinger. La disidencia de este último es muchísimo menos conocida. Su desafío al atomismo establecido prácticamente no ha trascendido.

Pero todo ello creemos que no es más que un aspecto superficial de una cuestión más profunda, y que podemos denominar, un tanto pomposamente, *la conversión de un fracaso en un éxito.* Así, la crisis en que se sumió la teoría cuántica antigua (y la Física en general) a principios de los años 20 del siglo pasado, y que conllevó prescindir de la visualización y la causalidad en el reino atómico, no se quiso ni se quiere llevar hasta el final. Seguimos haciéndonos imágenes del mundo microscópico, a pesar de que hace tiempo que se ha descubierto que ello es falsificador. Si bien, como hemos visto, hay que distinguir entre electrones ligados a un átomo y electrones libres, tampoco este último caso está exento de problemas, pues una partícula libre, por definición, no es observable. Y de hecho raramente se llega a un nivel de finura que haga distingos precisos en ese aspecto. Así, la complementariedad de Bohr no sería más que el intento filosófico de convertir en explicación lo que originalmente fue el colapso de un modelo explicativo fuertemente enraizado en la cosmovisión occidental.

Los legos en la materia, al vislumbrar contradicciones en el misterioso terreno cuántico, asumen humildemente su ignorancia ante enigmas cuya resolución está reservada a la elite de los elegidos. Pero los expertos dan cuenta de esas y otras paradojas como si se trataran de meras consecuencias de un mundo maravilloso, casi onírico, cuando -como dijo Schrödinger en repetidas ocasiones- las más de las veces las deslumbrantes aporías no son sino resultado de imágenes engañosas que se utilizan de manera injustificada y contraria a los desvelos de aquellos años fundacionales.

## 6. Referencias

**Agradecimientos**